\documentclass[10pt, twocolumn, letterpaper, journal]{IEEEtran}

\usepackage{amssymb,amsmath,epsfig,tabularx,delarray,enumerate,subfigure,graphicx,array,cite,mathdots,color}

\begin{document}
\title{Biosensing-by-learning Direct Targeting Strategy for Enhanced Tumor Sensitization}
\author{Yifan Chen$^\ast$, Muhammad Ali, Shaolong Shi, and U Kei Cheang
\thanks{\emph{Asterisk indicates corresponding author.}}
\thanks{$^\ast$Y. Chen and M. Ali are with the Faculty
of Computing and Mathematical Sciences, the University of Waikato,
Hamilton, New Zealand (e-mail: yifan.chen@waikato.ac.nz).}\thanks{S.
Shi is with Harbin Institute of Technology, Harbin, China. He is
also with the Department of Computer Science and Engineering,
Southern University of Science and Technology, Shenzhen,
China.}\thanks{U. K. Cheang is with the Department of Mechanical and
Energy Engineering, Southern University of Science and Technology,
Shenzhen, China.}}

\markboth{Submitted to IEEE Transactions on Nanobioscience}{Shell
\MakeLowercase{\textit{et al.}}: Biosensing-by-learning Direct
Targeting Strategy for Enhanced Tumor Sensitization}

\maketitle
\begin{abstract}
\emph{Objective:} We propose a novel iterative-optimization-inspired
direct targeting strategy (DTS) for smart nanosystems, which harness
swarms of externally manipulable nanoswimmers assembled by magnetic
nanoparticles (MNPs) for knowledge-aided tumor sensitization and
targeting. We aim to demonstrate through computational experiments
that the proposed DTS can significantly enhance the accumulation of
MNPs in the tumor site, which serve as a contrast agent in various
medical imaging modalities, by using the shortest possible
physiological routes and with minimal systemic exposure.

\emph{Methods:} The epicenter of a tumor corresponds to the global
maximum of an externally measurable objective function associated
with an \emph{in vivo} tumor-triggered biophysical gradient; the
domain of the objective function is the tissue region at a high risk
of malignancy; swarms of externally controllable magnetic
nanoswimmers for tumor sensitization are modeled as the guess
inputs. The objective function may be resulted from a passive
phenomenon such as reduced blood flow or increased kurtosis of
microvasculature due to tumor angiogenesis; otherwise, the objective
function may involve an active phenomenon such as the fibrin formed
during the coagulation cascade activated by tumor-targeted
``activator'' nanoparticles. Subsequently, the DTS can be
interpreted from the iterative optimization perspective: guess
inputs (i.e., swarms of nanoswimmers) are continuously updated
according to the gradient of the objective function in order to find
the optimum (i.e., tumor) by moving through the domain (i.e., tissue
under screening). Along this line of thought, we propose the
computational model based on the gradient descent (GD) iterative
method to describe the GD-inspired DTS, which takes into account the
realistic \emph{in vivo} propagation scenario of nanoswimmers.

\emph{Results:} By means of computational experiments, we show that
the GD-inspired DTS yields higher probabilities of tumor
sensitization and more significant dose accumulation compared to the
``brute-force" search, which corresponds to the systemic targeting
scenario where drug nanoparticles attempt to target a tumor by
enumerating all possible pathways in the complex vascular network.

\emph{Conclusion:} The knowledge-aided DTS has potential to enhance
the tumor sensitization and targeting performance remarkably by
exploiting the externally measurable, tumor-triggered biophysical
gradients.

\emph{Significance:} We believe that this work motivates a novel
biosensing-by-learning framework facilitated by externally
manipulable, smart nanosystems.
\end{abstract}


\section{Introduction}
\subsection{Background}
\subsubsection{Contrast-enhanced Medical Imaging}
Magnetic resonance imaging (MRI) is one of the standard procedures
for non-invasive clinical diagnosis of cancers due to its high soft
tissue contrast, spatial resolution, and penetration depth
\cite{TT17}. In addition, images are acquired without the use of
ionizing radiation or radio tracers that would cause harmful
side-effects. Contrast agents such as magnetic nanoparticles (MNPs)
are commonly used in MRI to provide better delineation between
healthy and diseased tissues \cite{TT17}. Another promising modality
for cancer diagnosis and recurrence monitoring is microwave imaging
in view of its safety, mobility, and cost-effectiveness
\cite{MGR17}. For example, a number of operational microwave breast
imaging systems are already in clinical use
\cite{YSH17,SSK17,PCP16,PCS16,MKM13} as reviewed in \cite{OOM18}. A
major challenge faced by this approach is the potentially small
dielectric contrast between tumor and its surrounding tissues, and
between benign and cancerous changes \cite{LPM07,OLB07,SEM09}. To
overcome these issues, MNPs have also been proposed as a contrast
agent \cite{BBB17,BCS15}. However, the current systemic targeted
drug delivery route can only deliver a very small fraction ($<2\%$)
of the administered nanoparticles to the precise site \cite{BP11}.
The main constraints include the reliance on systemic circulation,
the lack of a propelling force, and the absence of a sensory-based
displacement capability \cite{FMT16}.

\subsubsection{Amplification of Tumor Homing through Externally Manipulable Nanoswimmers}
Enhancing the diagnostic efficacy of contrast agents necessitates
the use of a direct targeting strategy (DTS) that allows agents to
reach the target tissues using the shortest physiological routes and
with minimal systemic exposure. In \cite{FMT16}, swarms of
magneto-aerotactic bacteria, namely Magnetococcus marinus strain
MC-1, are harnessed for delivering drug-containing nanoliposomes to
the disease site to improve the therapeutic index of various
nanocarriers in tumor regions. MC-1 cells, each containing a chain
of magnetic iron oxide nanocrystals, tend to swim along local
magnetic field lines and towards low oxygen concentrations based on
a two-state aerotactic sensing system. It was shown that when MC-1
cells were injected near the tumor and magnetically guided, up to
$55\%$ of MC-1 cells penetrated into hypoxic regions of the tumor.
Furthermore, nanoswimmers assembled by MNPs have also been proposed
for direct targeting, which use magnetic self-assembly of
$50-100~\mathrm{nm}$ iron oxide nanoparticles \cite{CK15}. Under an
external magnetic field, the MNPs can magnetize and form chains that
are flexible under time-varying magnetic fields \emph{via}
magnetohydrodynamics. A coil system was designed to actuate the
nanoswimmers by applying a nearly uniform magnetic field through the
Helmholtz configuration \cite{CMK16,CKM17}. One common external
force can control large numbers of nanoswimmers to perform a complex
task such as penetration of a tumor cell membrane for the selective
release of a drug inside the cell \cite{CMK16,MER18}. However,
nanoswimmers-assisted direct targeting of contrast agents requires
\emph{a priori} knowledge about the location of the disease site,
which is usually unavailable if the image quality is too low in the
pre-contrast medical imaging. This results in a chicken-or-egg
dilemma.

\subsubsection{Amplification of Tumor Homing through Smart Nanosystems}
Another strategy to amplify disease targeting is to design smart
nanosystems that leverage the living host environment
\cite{SNM18,OZ18,KLO17,ZWS16,KLB15,VPL11,PVO10,ASK10,PVX10,SDP07}.
These nanosystems can be classified in two categories:
environment-responsive and environment-primed \cite{KLB15}. The
former category encompasses nanoparticles that sense and
subsequently respond to their environment. Altered \emph{in vivo}
conditions such as redox potential, pH, enzymatic activity, and
homeostatic pathways (see Fig. 1 in \cite{KLB15} for a comprehensive
overview of various mechanisms) induced by disease conditions can be
leveraged to mobilize nanoparticle systems that are administered in
these preexisting contexts. The latter category is defined by an
emerging paradigm of cooperative nanosystems, such that the host
environment is manipulated by an external influence to enable
desired host-nanoparticle and nanoparticle-nanoparticle
interactions, such as communication, recruitment, or amplification.
Modifications to the host that achieve this primed environment can
be accomplished by administering energy (X-rays, infrared light,
heat), drugs, or nanoparticles themselves. For example, the
nanosystem in \cite{PVX10} consists of two components. The first
component is gold nanorods that populate the porous tumor vessels
\emph{via} systemic targeting by utilizing the conventional enhanced
permeability and retention (EPR) effect and then act as photothermal
antennas to specify tumor heating \emph{via} remote near-infrared
laser irradiation. Local tumor heating accelerates the recruitment
of the second component: a targeted nanoparticle consisting of
either a prototypical imaging agent (magnetofluorescent iron oxide
nanoworms) or a prototypical therapeutic agent (doxorubicin-loaded
liposomes). In \cite{VPL11}, gold nanorods or engineered proteins
target tumors and then locally activate the coagulation cascade to
broadcast tumor location to clot-targeted nanoworms or liposomes in
circulation. Smart nanosystems do not require location information
of the disease site. However, they still rely on systemic
circulation for homing to cancer cells without using an external
guidance.
\begin{figure} [!htp]
\begin{center}
\epsfig{file=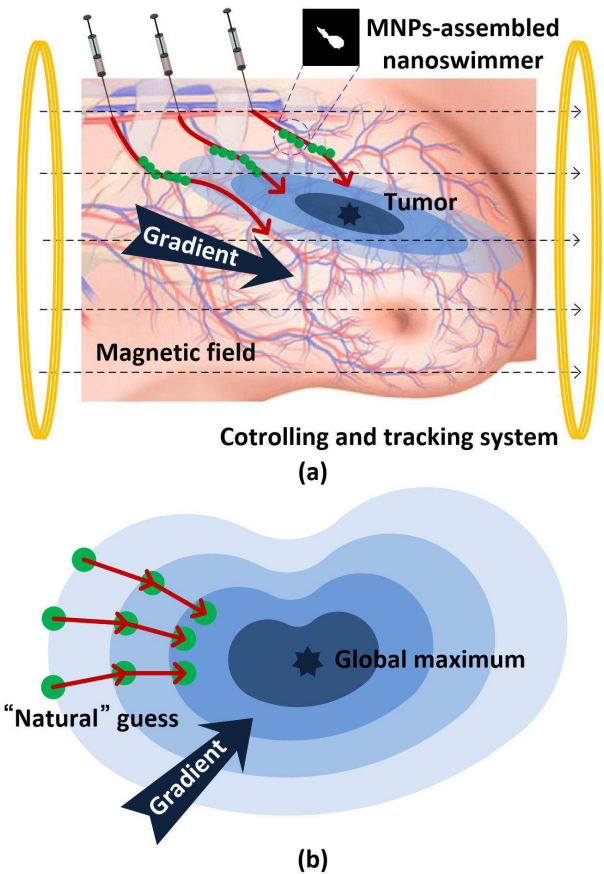,width=0.9\linewidth}
\end{center}
    \caption{Analogy between (a) DTS employed in an externally manipulable,
    smart nanosystem for tumor sensitization, and (b) iterative optimization process.}
    \label{fig:Fig1}
\end{figure}

\subsection{Biosensing by Learning}
The aforementioned experimental investigations provide the basis for
an externally manipulable, smart nanosystem, where non-manipulable
nanoparticles can be replaced by nanoswimmers assembled by iron
oxide MNPs \cite{CK15}, as depicted in Fig. 1(a). The magnetic
response of MNPs, induced by a polarizing magnetic field, allows for
reliable estimation of the locations of magnetic changes through
differential medical imaging \cite{BBB17,BCS15}. For
environment-responsive operations, an external controlling and
tracking system probes the host environment by analyzing the
measurable characteristics of nanoswimmers (e.g., trajectories,
magnetic changes induced) and steers them towards the direction
where a tumor is likely to be present as shown in Fig. 1(a). For
environment-primed operations, gold nanorods first prime the host
environment \cite{PVO10,PVX10,VPL11} to interact with nanoswimmers.
Similarly, the external system monitors the \emph{in vivo} responses
by observing the properties of nanoswimmers and maneuver them
correspondingly to enhance tumor sensitization. Should a specific
tissue region be a tumor, MNPs will accumulate in the region on the
basis of the EPR effect or receptor-ligand binding, which can be
observed externally by noticing that MNPs appear to stop moving
\cite{CSY17,CNK16,CKM13}.

The current investigation attempts to develop a computational model
for feasibility study of the proposed intelligent (i.e.,
knowledge-aided) DTS. Nature's blueprints have inspired exciting new
fields of science such as bio-inspired computing that creates
problem-solving techniques using insights from natural systems. For
example, the process of natural selection inspired the development
of the classical genetic algorithm to solve complex optimization and
search problems. It is also stimulating to look the other way by
exploiting computing strategies for biomedical applications
\cite{CNK16,CSY17}. There is an intriguing analogy between the
knowledge-aided DTS in an externally manipulable nanosystem for
tumor sensitization (Fig. 1(a)) and the iterative optimization
process (Fig. 1(b)). The global maximum of a unimodal, externally
measurable objective function corresponding to a tumor-induced
biophysical phenomenon is the tumor to be detected; the domain of
the function is the tissue region at a high risk of malignancy; the
guess solution is a swarm of externally manoeuvrable magnetic
nanoswimmers. A guess input (i.e., nanoswimmers) locates the optimal
solution (i.e., cancer) by moving through the domain (i.e.,
high-risk tissue) under the guidance of a specified force (i.e.,
steering field). The objective function may be altered by the guess
made of natural materials because the guess input interacts with the
domain (i.e., nanoswimmers undergo physical, chemical, and
biological interactions with the host environment). This is in
contrast to a traditional iterative method using a non-interacting
approximate solution. An external observer can then infer the domain
by monitoring the movement of the guess (``seeing-is-sensing''
\cite{CNK16}), where the $(n+1)^\mathrm{th}$ approximation is
derived from the $n^\mathrm{th}$ one. This strategy is within the
general framework of computing-inspired bio-detection proposed in
our previous work \cite{CSY17}. Provided with the analogy, a wide
variety of iterative methods can thus be applied to the design of an
optimal DTS. To elaborate on the proposed methodology, the classical
gradient descent (GD) method is used to inspire the DTS, where the
guess input takes steps based on the gradient of the objective
function at the current point. Furthermore, the derivative of the
function needs to be approximated in real-time and the movement of
the guess is constrained by the physical conditions of human
microvasculature.

It is worth noting that, from the computational perspective, the
traditional systemic delivery of contrast agents can be regarded as
a ``brute-force" search where contrast agent nanoparticles attempt
to detect a tumor \emph{via} a medical imaging system by enumerating
all possible pathways in the complex vascular network and checking
whether each pathway is intercepted by a tumor. Furthermore, the
original smart nanosystems in
\cite{SNM18,OZ18,KLO17,ZWS16,KLB15,VPL11,PVO10,ASK10,PVX10,SDP07}
can be regarded as a brute-force search given an \emph{expanded}
tumoral region due to tumor target amplification facilitated by the
peritumoral biophysical conditions (for environment-responsive
nanosystems) or the initial-stage triggering modules (for
environment-primed nanosystems).

\subsection{Organization of the Paper}
The paper is organized as follows. In Section II, we discuss the
suitable \emph{in vivo} biophysical gradients that can be mapped to
externally measurable objective functions. In Section III, we
analyze the propagation model of nanoswimmers in a discretized
capillary network, followed by the discussion on the general
iterative DTS framework and some representative functions showing
different situations that the DTS has to face. In Section IV, we
propose the GD-inspired DTS subject to the realistic physical
constraints of controlling and tracking nanoswimmers \emph{in vivo}.
In Section V, we provide numerical examples to demonstrate the
effectiveness of the proposed framework. Finally, some concluding
remarks are drawn in Section VI.

\section{Externally Measurable Objective Functions}
In the current work, tumor sensitization is performed indirectly
through an external controlling and tracking system as shown in Fig.
1(a), such as an integrated device consisting of multiple pairs of
electromagnetic coils to generate the rotating magnetic field to
actuate the magnetic nanoswimmers \cite{CMK16,CKM17} and another
coil to supply the polarizing magnetic field inducing the magnetic
contrast associated with the nanoswimmers \cite{BBB17,BCS15}.
Therefore, it is necessary that the \emph{in vivo} biophysical
gradients can be mapped to an externally measurable objective
function by using nanoswimmers as a probe for analysis of the host
environment.

\subsection{Environment-responsive Nanosystems}
For this type of nanosystems, passive physical properties of the
host environment such as peritumoral vascular architecture
\cite{GBL95,BGB96,BJ00} and blood flow velocity
\cite{FJ07,WII09,BGB96,KKL09} can be exploited to derive the
biological gradients. Oxygen and nutrients are supplied to cancer
cells \emph{via} new blood vessels that have extended into the
cancer tissue. Typical skeletonized images of various classes of
vascular networks demonstrate that normal capillaries exhibit almost
uniformly distributed grid patterns to ensure adequate oxygen
transportation throughout the tissue \cite{GBL95,BGB96,BJ00}. On the
other hand, tumor vessels have a profound sort of tortuosity with
many smaller bends on each larger bend \cite{GBL95,BGB96,BJ00}. In
terms of blood flow velocity, its value in tumor tissues is
significantly lower than that in healthy tissues due to the
hypovascular structure of the malignant lesion \cite{KKL09,FJ07}.
This phenomenon has been observed for cancer cells in the visceral
pleura \cite{WII09}, malignant gliomas \cite{BGB96}, and pancreatic
tumors \cite{KKL09}. In summary, the externally measurable objective
functions corresponding to the aforementioned two biophysical
conditions can be derived from the variations in the tortuosity of
nanoswimmer trajectory and the resultant nanoswimmer velocity,
respectively, with respect to the values for normal tissues. Both of
them would increase as the distance between the nanoswimmer and the
tumor decreases.

\subsection{Environment-primed Nanosystems}
For this type of systems, specific ``activator'' nanomaterials can
be used to detect a diseased site and act as tumor-specific triggers
to induce biophysical gradients. For example, gold nanorods can be
modified to circulate for long periods of time in the blood stream
and be passively accumulated in tumors \emph{via} systemic
circulation \cite{PVO10,PVX10,VPL11}. They are used to heat tumor
tissues by amplifying the absorption of near-infrared energy that is
mostly transparent to living tissues \cite{PVO10,PVX10,VPL11}. The
associated photothermal heating is highly localized around the tumor
site. Consequently, the gradient of blood flow velocity is amplified
due to the differential response of normal and tumor
microcirculation to hyperthermia, where blood flow in normal tissue
increases much faster with temperature and stasis occurs at higher
levels of hyperthermia compared to tumors owning to the rapid growth
of tumor cell population relative to deteriorating vascular beds
\cite{DJ84,SON84}. Furthermore, local heating disrupts tumor vessels
and initiates extravascular coagulation. Hence, the fibrin forms the
coagulation gradient centered at the tumor caused by temperature
increase. The magnetic nanoswimmers employ the peptide coatings that
recognize fibrin directly for clot targeting
\cite{PVO10,PVX10,VPL11}. Direct binding in regions of coagulation
will reduce the concentration of mobile nanoswimmers under tracking.
In summary, the externally measurable objective functions associated
with the aforementioned two phenomena can be derived from the
variations in the resultant nanoswimmer velocity and the magnetic
contrast induced by nanoswimmers, respectively, with respect to the
values for normal tissues. Both of them would increase as the
distance between the nanoswimmer and the tumor decreases. In
addition, local hyperthermia results in a temperature gradient from
the tissue malignancy to its peripheral region, which may be
directly measured from the infrared thermographic imaging if the
tumor is close to the skin \cite{PVO10,PVX10,VPL11}. In this case,
the global gradient towards the tumor epicenter can be readily
obtained.

\section{Iterative DTS}
\subsection{Invasion-percolation-based Multilayer Vascular Network Model}
Tumor vasculature is more chaotic in appearance than normal
vasculature, which can be measured using fractal geometries
\cite{BJ00}. For example, tumor vessels yield fractal dimensions of
$1.89\pm0.04$, whereas normal arteries and veins yield dimensions of
$1.70\pm0.03$, and normal capillaries produce essentially
two-dimensional patterns \cite{GBL95,BGB96,BJ00}. It was also
observed that the microvascular density in the peritumoral region
increases due to the supply of growth factors from the tumor and
reduces in the tumor center due to a combination of severely reduced
blood flow and solid stress exerted by the tumor \cite{LRB06}.

Consequently, it is assumed that normal tissues are regularly
vascularized, which results in a homogeneous lattice comprised of
straight, rigid cylindrical capillaries that join adjacent nodes
\cite{MAC02,BGB96}. On the other hand, the observed fractal
dimensions of tumor vasculature can be described by the invasion
percolation process \cite{GBL95,BGB96,BJ00}, which is implemented by
first assigning uniformly distributed random values of strengths to
each point on the underlying square lattice representing potential
paths of vascular growth. Starting at an arbitrary site the network
occupies the lattice point adjacent to the current site that has the
lowest strength. Growth is iterated until the desired lattice
occupancy is reached. Blood vessels are assumed to connect all
adjacent occupied lattice points. \begin{figure} [!htp]
\begin{center}
\epsfig{file=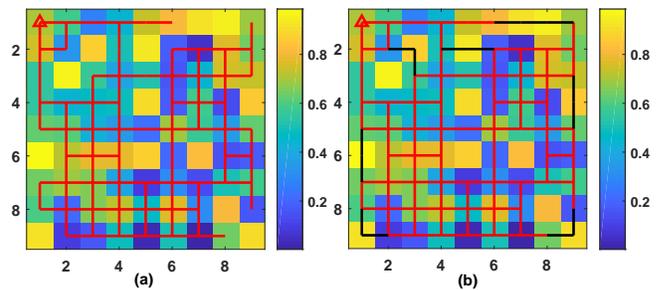,width=\linewidth}
\end{center}
    \caption{(a) An invasion percolation network after 100 growth
    steps, and (b) adding vessels to ensure nonzero blood flow throughout the
    network.}
    \label{fig:Fig2}
\end{figure}\begin{figure} [!htp]
\begin{center}
\epsfig{file=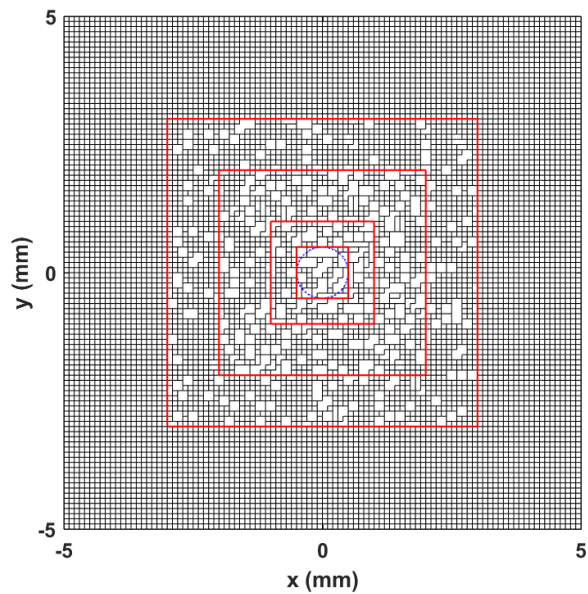,width=\linewidth}
\end{center}
    \caption{Simulated multi-layer vascular network. The
    level of occupancy on the lattice reduces from $100\%$ to $40\%$
    gradually as the distance to the tumor center decreases. The
    boundaries of the layers are denoted by the red solid lines.
    The tumor center is denoted by the blue dotted circle.}
    \label{fig:Fig3}
\end{figure}Finally, additional edges are added to
``pathological'' nodes to ensure nonzero blood flow throughout the
entire percolation cluster. The simulated networks may be matched
with real tumor vasculature by selecting appropriate occupancy
levels. Following \cite{BGB96}, the fractal dimensions are around
$1.6$, $1.8$, $1.9$, and $2.0$ for $40$, $60$, $80$, and $100\%$
occupancy on the backbone, respectively.

Moreover, malignant tumors often possess fuzzy and blurred
boundaries \cite{REL97,SSB06}. As such, the fractal dimensions
across the boundary of a tumor can be characterized by a smooth
transition from inside a tumor to the outside. To quantify the
diffusive nature of a tissue anomaly, a discretized multilayer model
can be applied to approximate the gradual, continuous change in the
fractal dimension across the periphery of a lesion. Fig. 2(a)
depicts an invasion percolation network after $100$ growth steps,
and Fig. 2(b) shows additional vessels to ensure nonzero blood flow
throughout the network assuming that the blood inflow and outflow
are in the top left and bottom right. \begin{figure} [!htp]
\begin{center}
\epsfig{file=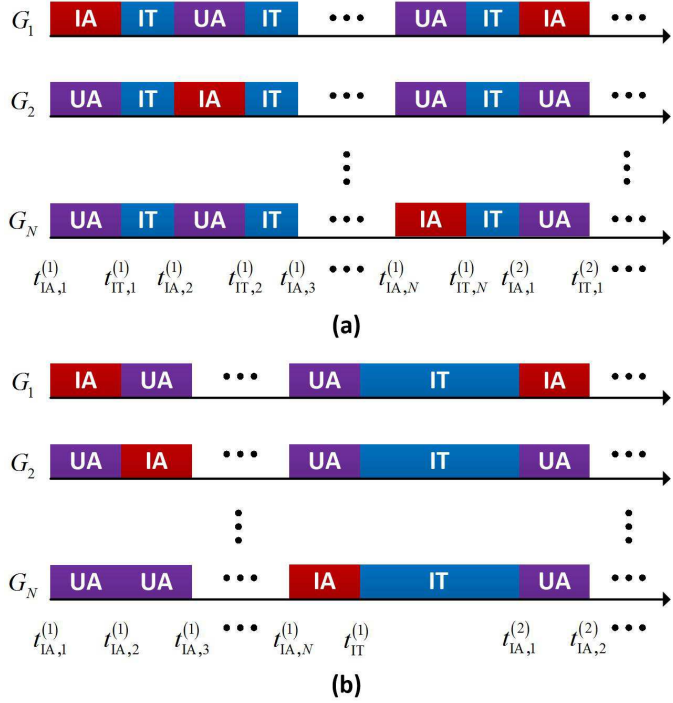,width=\linewidth}
\end{center}
    \caption{Time sequence of system operational modes in one cycle for multiple guess inputs. Each guess has three
    operational modes: intended actuating (IA), imaging and tracking (IT), and
    unintended actuating (UA). Two time-division multiplexing (TDM) protocols are considered. (a) TDM1: each guess
    takes turn to operate in the IA and IT modes; and (b) TDM2: each guess takes turn to operate in the IA mode, followed by a common
    IT mode.}
    \label{fig:Fig4}
\end{figure}Fig. 3 illustrates a simulated
multilayer vascular network, where the intercapillary distance is
set to be $100~\mu m$, and the level of occupancy on the lattice
reduces from $100\%$ to $40\%$ gradually as the distance to the
tumor center decreases.

\subsection{Problem Formulation}
Let $f$ represent an externally measurable objective function and be
defined on the domain $\mathbb{D}$, which denotes the high-risk
tissue region under surveillance. The landscape of $f$ is distorted
by a ``natural'' guess $G$ representing a swarm of magnetic
nanoswimmers as follows
\begin{equation} 
\begin{split}
f\left(\vec{x};G\right)&=f_\mathrm{A}\left(\vec{x};G\right)+f_\mathrm{C}\left(\vec{x};G\right)\\
&=f_\mathrm{T}(\vec{x})+f_{\mathrm{D}}\left(\vec{x};G\right)+f_\mathrm{C}\left(\vec{x};G\right),~~~\vec{x}\in\mathbb{D},
\end{split}
\end{equation}
where $f_\mathrm{A}\left(\vec{x};G\right)$ is the \emph{apparent}
objective function measured at location $\vec{x}$ through guess $G$,
$f_\mathrm{T}(\vec{x})$ is the \emph{true} objective function at
$\vec{x}$ independent of the presence or absence of $G$,
$f_{\mathrm{D}}\left(\vec{x};G\right)$ is the disturbance resulted
from the interaction between $G$ and the domain $\mathbb{D}$, and
$f_\mathrm{C}\left(\vec{x};G\right)$ is the correction factor
accounting for the disturbance caused by $G$. For a meaningful
optimization process, it is assumed that regardless of any variation
caused by the guess to the function, the location of the global
maximum denoting the tumor, $\vec{x}^\star$, remains unchanged.

The true objective $f_\mathrm{T}(\vec{x})$, dependent on the
underlying tumor-triggered biophysical phenomena, may take the form
of variation in path tortuosity, velocity, or magnetic contrast of
nanoswimmers as discussed earlier. Subsequently, for the measure of
tortuosity, an alteration $f_{\mathrm{D}}(\vec{x};G)$ would incur if
the nanoswimmers are engineered to modify the vasculature of tumors
(e.g., anti-angiogenic agents to shut down tumor vessels or
pro-angiogenic agents to normalize tumor vessels) \cite{KLB15}. For
the measure of velocity, $f_{\mathrm{D}}(\vec{x};G)$ is given by the
relative velocity of nanoswimmers with respect to the blood stream.
In the case of magnetic change, $f_{\mathrm{D}}(\vec{x};G)$ is
proportional to the reduction in the concentration of nanoswimmers
due to various loss mechanisms such as degradation (nanoswimmers
degenerate in the blood), branching (nanoswimmers move into an
unintended vascular branch), and diffusion (random motions of
nanoswimmers driven by the concentration gradient) \cite{CKA15}.
Finally, the correction factor $f_\mathrm{C}\left(\vec{x};G\right)$
attempts to counteract $f_{\mathrm{D}}\left(\vec{x};G\right)$ to
minimize its influence on the true landscape, i.e.,
$f_\mathrm{C}\left(\vec{x};G\right)=-f_{\mathrm{D}}\left(\vec{x};G\right)+\chi\left(\vec{x};G\right)$
with $\chi\left(\vec{x};G\right)$ being the random compensation
error. Therefore, Eq. (1) can be rewritten as
\begin{equation} 
\begin{split}
f\left(\vec{x};G\right)=f_\mathrm{T}(\vec{x})+\chi\left(\vec{x};G\right),~~~\vec{x}\in\mathbb{D}.
\end{split}
\end{equation}

In the computational framework of DTS, multiple guess inputs
$G_1,G_2,\cdots,G_N$ are first deployed in multiple pre-specified
sites
$\mathbb{R}_1,\mathbb{R}_2,\cdots,\mathbb{R}_N\subseteq\mathbb{D}$,
where $\mathbb{R}_n~(n=1,2,\cdots,N)$ denote the injection sites of
nanoswimmers as depicted in Fig. 1. The guesses begin searching for
the optimal solution following some iterative algorithms. The DTS
includes the following key steps.
\begin{enumerate}
    \item \emph{Initialization.} The guess inputs are
    deployed in $\mathbb{R}_1,\mathbb{R}_2,\cdots,\mathbb{R}_N$ at
    the same starting times $t^{(1)}_{\mathrm{IA},1}$ as
    shown in Fig. 4(a)-(b) with initial locations
    $\vec{x}_1\left(t^{(1)}_{\mathrm{IA},1}\right),\vec{x}_2\left(t^{(1)}_{\mathrm{IA},1}\right),
    \cdots,\vec{x}_N\left(t^{(1)}_{\mathrm{IA},1}\right)$, respectively. Suppose that the external system
    operates in the simple time-multiplexed manner. Without loss of generality, consider the
    first guess input $G_1$, which operates on the following three
    modes: Intended Actuating (IA), Imaging and Tracking (IT), and
    Unintended Actuating (UA). Two time-division multiplexing (TDM) protocols are considered. For TDM1 each guess
    takes turn to operate in the IA and IT modes, whereas for TDM2 each guess takes turn to operate in the IA mode, followed by a common
    IT mode as illustrated in Fig. 4(a) and 4(b), respectively.
    \item \emph{IA.} For TDM1, from
    $t^{(1)}_{\mathrm{IA},1}$ to $t^{(1)}_{\mathrm{IT},1}$, $G_1$ operates in the IA mode and its
    trajectory is determined by the angle deviation relative to a
    principal axis denoting an intended steering vector upon $G_1$ at
    $\vec{x}_1\left(t^{(1)}_{\mathrm{IA},1}\right)$, $\phi\left(t^{(1)}_{\mathrm{IA},1}\right)$,
    which indicates a uniform magnetic field in the surveillance
    domain \cite{CMK16,CKM17} and is dependent on the iterative method described in Section IV.
    The next location of $G_1$ at time instant
    $t^{(1)}_{\mathrm{IT},1}$ is then updated according to:\begin{figure} [!htp]
\begin{center}
\epsfig{file=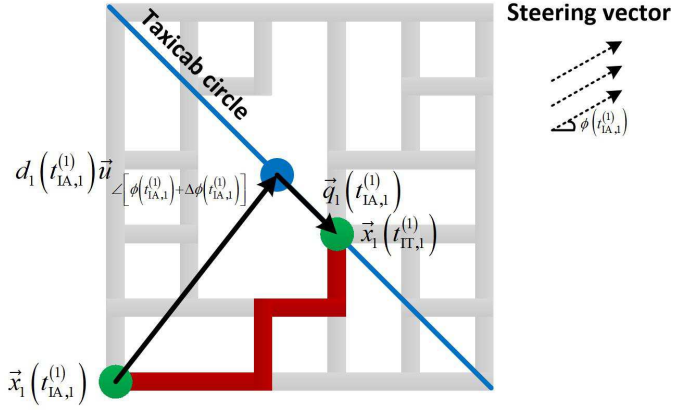,width=\linewidth}
\end{center}
    \caption{Updating of a guess input in the taxicab vascular network subject to a
    specified steering vector.}
    \label{fig:Fig5}
\end{figure}
    \begin{equation} 
    \begin{split}
    &\vec{x}_1\left(t^{(1)}_{\mathrm{IT},1}\right)\\
    &=\vec{x}_1\left(t^{(1)}_{\mathrm{IA},1}\right)
    +d_1\left(t^{(1)}_{\mathrm{IA},1}\right)\vec{u}_{\angle\left[\phi\left(t^{(1)}_{\mathrm{IA},1}\right)
    +\Delta\phi\left(t^{(1)}_{\mathrm{IA},1}\right)\right]}\\
    &~~~+\vec{q}_1\left(t^{(1)}_{\mathrm{IA},1}\right).
    \end{split}
    \end{equation}
    The term $\vec{u}_{\angle\phi}$ denotes a unit vector with angle
    $\phi$ and $\Delta\phi\left(t^{(1)}_{\mathrm{IA},1}\right)$ is a
    random variable summarizing all steering imperfections, which is
    assumed to be normally distributed with variance
    $\sigma^2_{\Delta\phi}$ and zero mean for simplicity.
    The displacement length $d_1\left(t^{(1)}_{\mathrm{IA},1}\right)$
    satisfies
    \begin{equation} 
    \begin{split}
    &\left\|d_1\left(t^{(1)}_{\mathrm{IA},1}\right)\vec{u}_{\angle\left[\phi\left(t^{(1)}_{\mathrm{IA},1}\right)
    +\Delta\phi\left(t^{(1)}_{\mathrm{IA},1}\right)\right]}\right\|_1\\
    &=\bigg|d_1\left(t^{(1)}_{\mathrm{IA},1}\right)\cos\left[\phi\left(t^{(1)}_{\mathrm{IA},1}\right)+
    \Delta\phi\left(t^{(1)}_{\mathrm{IA},1}\right)\right]\bigg|\\
    &~~~+\bigg|d_1\left(t^{(1)}_{\mathrm{IA},1}\right)\sin\left[\phi\left(t^{(1)}_{\mathrm{IA},1}\right)+
    \Delta\phi\left(t^{(1)}_{\mathrm{IA},1}\right)\right]\bigg|\\
    &=v_1\left(t^{(1)}_{\mathrm{IA},1}\right)\left(t^{(1)}_{\mathrm{IT},1}-t^{(1)}_{\mathrm{IA},1}\right),
    \end{split}
    \end{equation}
    where $\|\cdot\|_1$ denotes the $\ell_1$ norm and $v_1\left(t^{(1)}_{\mathrm{IA},1}\right)$ is the velocity
    of $G_1$ at $t^{(1)}_{\mathrm{IA},1}$ given the taxicab geometry of the
    vascular network. Finally,
    $\vec{q}_1\left(t^{(1)}_{\mathrm{IA},1}\right)$ is the position
    ``quantization'' error due to the discrete lattice pattern of the
    vasculature as illustrated in Fig. 5, which is the displacement
    vector from the point $\vec{x}_1\left(t^{(1)}_{\mathrm{IA},1}\right)+
    d_1\left(t^{(1)}_{\mathrm{IA},1}\right)\vec{u}_{\angle\left[\phi\left(t^{(1)}_{\mathrm{IA},1}\right)
    +\Delta\phi\left(t^{(1)}_{\mathrm{IA},1}\right)\right]}$
    on the continuous taxicab circle of radius
    $v_1\left(t^{(1)}_{\mathrm{IA},1}\right)\left(t^{(1)}_{\mathrm{IT},1}-t^{(1)}_{\mathrm{IA},1}\right)$,
    to its closest point in the discrete vascular network having the same taxicab
    distance to $\vec{x}_1\left(t^{(1)}_{\mathrm{IA},1}\right)$.

    For TDM2, the same process as mentioned above applies except
    that $t^{(1)}_{\mathrm{IT},1}$ is replaced by
    $t^{(1)}_{\mathrm{IA},2}$ as shown in Fig. 4(b).

    \item \emph{IT.} For TDM1, from
    $t^{(1)}_{\mathrm{IT},1}$ to $t^{(1)}_{\mathrm{IA},2}$, $G_1$ operates in the IT mode.
    In the absence of a steering field, $G_1$ follows a random walk in the
    lattice (i.e., at each intersection it has the same probability to either move up or to the right)
    and swims towards various locations in equal observation time
    intervals along a zigzag pathway,
    $\vec{x}_1\left(t^{(1)}_{\mathrm{IT},1}\right)$,
    $\vec{x}_1\left(t^{(1)}_{\mathrm{IT},1}+\Delta t\right)$, $\cdots$, $\vec{x}_1\left(t^{(1)}_{\mathrm{IT},1}+K\Delta t\right)$,
    $\vec{x}_1\left(t^{(1)}_{\mathrm{IA},2}\right)$ as shown in Fig. 6, where $\Delta
    t=\left(t^{(1)}_{\mathrm{IA},2}-t^{(1)}_{\mathrm{IT},1}\right)/(K+1)$.
    Various imaging modalities such as MRI \cite{TT17} and microwave imaging
    \cite{BBB17} can be used to detect the
    magnetic contrast induced by multiple magnetic nanoswimmers simultaneously, which allows for
    tracking of all the nanoswimmers. In contrast to mathematical computing where the location of a
    guess input is known exactly, the guess location in the current
    ``natural'' computing needs to be estimated. The
    positioning error is summarized in the random variable
    $\Delta\vec{x}_1$ as also shown in Fig. 6, whose horizontal and
    vertical components are assumed to be independently
    and identically distributed Gaussian random variables with equal variance
    $\sigma^2_{\Delta x}$ and zero mean for simplicity.
    In that case, $\left|\Delta\vec{x}_1\right|$ is Rayleigh-distributed.
    The objective function is then evaluated at each location. For
    example, if the nanoswimmer velocity is considered, the values
    are obtained as
    \begin{equation} 
    \begin{split}
    &f\left(\vec{x}_1\left(t^{(1)}_{\mathrm{IT},1}+k\Delta t\right)
    +\Delta\vec{x}_1\left(t^{(1)}_{\mathrm{IT},1}+k\Delta t\right)\right)\\
    &\approx\frac{1}{\Delta t}\Big\|\vec{x}_1\left(t^{(1)}_{\mathrm{IT},1}+(k+1)\Delta
    t\right)\\
    &~~~~~~~~~+\Delta\vec{x}_1\left(t^{(1)}_{\mathrm{IT},1}+(k+1)\Delta t\right)\\
    &~~~~~~~~~-\vec{x}_1\left(t^{(1)}_{\mathrm{IT},1}+k\Delta
    t\right)\\
    &~~~~~~~~~-\Delta\vec{x}_1\left(t^{(1)}_{\mathrm{IT},1}+k\Delta
    t\right)\Big\|_1,~k=0,1,\cdots,K.
    \end{split}
    \end{equation}
    Subsequently, the gradient for guess $G_1$ at $t^{(1)}_{\mathrm{IT},1}$ is estimated as
    follows
    \begin{equation} 
    \begin{split}
    &\nabla f\left(\vec{x}_1\left(t^{(1)}_{\mathrm{IT},1}\right)\right)\\
    &\approx\max_{k_1,k_2}\\
    &\bigg\{\Big[f\left(\vec{x}_1\left(t^{(1)}_{\mathrm{IT},1}+k_1\Delta t\right)
    +\Delta\vec{x}_1\left(t^{(1)}_{\mathrm{IT},1}+k_1\Delta
    t\right)\right)\\
    &~~~-f\left(\vec{x}_1\left(t^{(1)}_{\mathrm{IT},1}+k_2\Delta t\right)
    +\Delta\vec{x}_1\left(t^{(1)}_{\mathrm{IT},1}+k_2\Delta
    t\right)\right)\Big]\bigg/\\
    &~~\Big\|\vec{x}_1\left(t^{(1)}_{\mathrm{IT},1}+k_1\Delta t\right)
    +\Delta\vec{x}_1\left(t^{(1)}_{\mathrm{IT},1}+k_1\Delta t\right)\\
    &~~~-\vec{x}_1\left(t^{(1)}_{\mathrm{IT},1}+k_2\Delta t\right)
    -\Delta\vec{x}_1\left(t^{(1)}_{\mathrm{IT},1}+k_2\Delta
    t\right)\Big\|_2
    \bigg\},\\
    &~~~~~~~~~~~~~~~~~~~~~~~k_1>k_2~\mathrm{and}~k_1,k_2\in\{0,1,\cdots,K\},
    \end{split}
    \end{equation}\begin{figure} [!htp]
\begin{center}
\epsfig{file=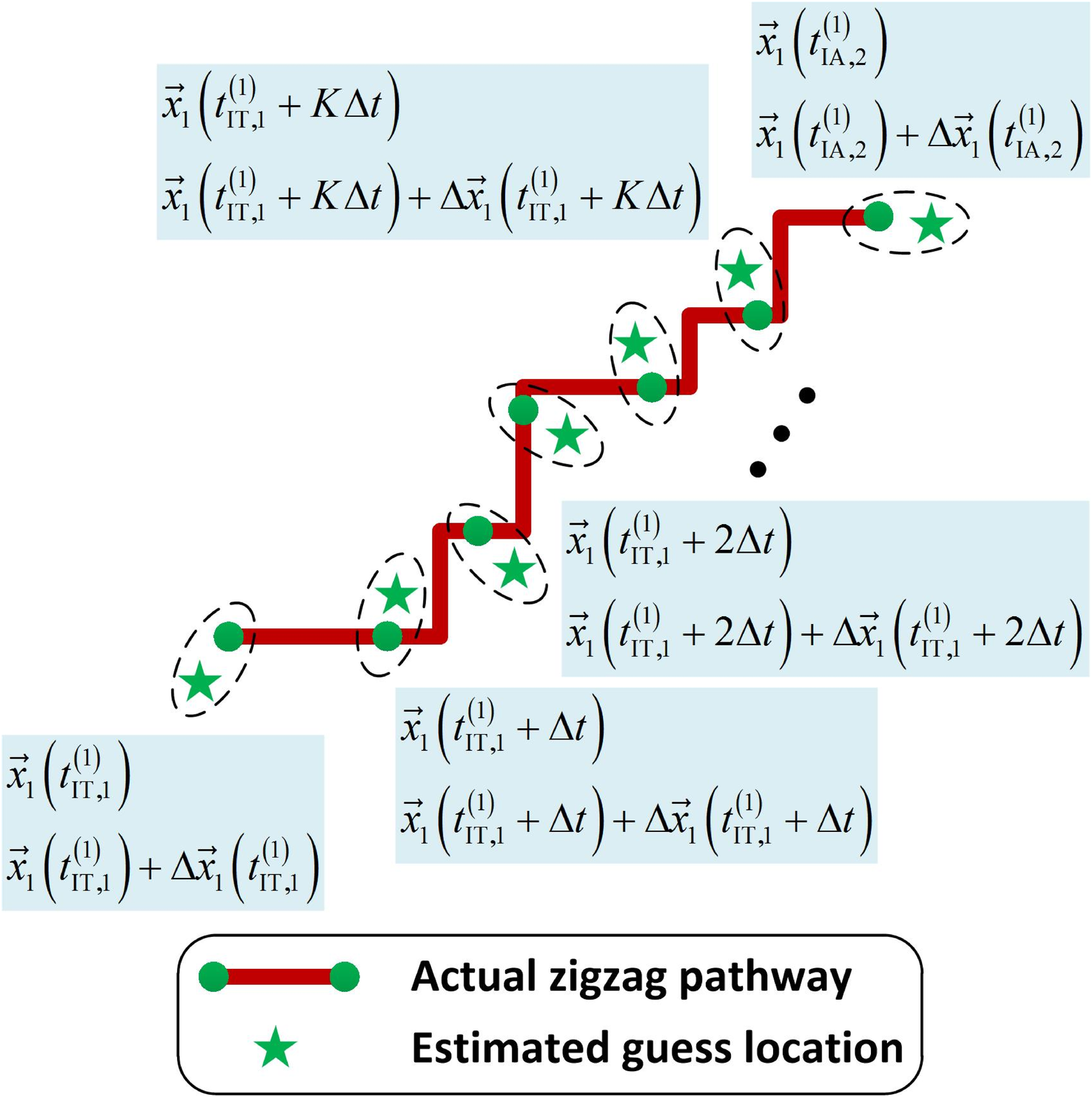,width=0.9\linewidth}
\end{center}
    \caption{Pictorial illustration of the IT process:
    the guess follows a random walk in the
    lattice and moves towards various locations along a zigzag pathway,
    $\vec{x}_1\left(t_{\mathrm{IT},1}\right)$,
    $\vec{x}_1\left(t_{\mathrm{IT},1}+\Delta t\right)$, $\cdots$, $\vec{x}_1\left(t_{\mathrm{IT},1}+K\Delta t\right)$,
    $\vec{x}_1\left(t_{\mathrm{IA},2}\right)$ with $\Delta t$ being the observation time interval.}
    \label{fig:Fig6}
\end{figure}
    where $\|\cdot\|_2$ is the $\ell_2$ norm. The overall gradient
    after $N$ IT processes is estimated by taking into account all
    the $N$ gradients obtained at $t^{(1)}_{\mathrm{IT},1},
    t^{(1)}_{\mathrm{IT},2},\cdots,t^{(1)}_{\mathrm{IT},N}$,
    respectively.
    A new steering vector for $G_1$ is then computed by following a
    specified algorithm as discussed in Section IV, which is
    used to guide the movement of $G_1$ during the next IA operation
    at $t^{(2)}_{\mathrm{IA},1}$ as shown in Fig. 4(a).
    As the nanoswimmer is in the form of nanochains or bundle-like aggregates
    assembled by MNPs \cite{CK15}, it has a finite lifespan due to the
    dissembling and diffusion of MNPs during propagation.
    In the case that $G_1$ is fully consumed in
    $\mathbb{D}$, a new guess input is deployed at $\mathbb{R}$.

    For TDM2, $G_1$ operates in the UA mode (as explained below) from $t^{(1)}_{\mathrm{IA},2}$
    to $t^{(1)}_{\mathrm{IA},3}$ when $G_2$ is in the IA mode.

    \item \emph{UA.} For TDM1, from
    $t^{(1)}_{\mathrm{IA},2}$ to $t^{(1)}_{\mathrm{IT},2}$, $G_1$ operates in the UA
    mode. This is similar to the IA operation except that the
    steering field is meant for the second guess $G_2$. This is due
    to the limitation of the current coil system in generating the
    steering field, which exerts a global uniform torque on all the
    nanoswimmers simultaneously instead of localized torques on
    individual nanoswimmers. The same DTS steps
    (i.e., $\mathrm{IA}\rightarrow\mathrm{IT}\rightarrow\mathrm{UA}$)
    are applied to all the guesses in sequence and the iteration
    continues unless certain stopping criteria are met.

    For TDM2, from $t^{(1)}_{\mathrm{IA},3}$
    to $t^{(1)}_{\mathrm{IA},4}$, $G_1$ again operates in the UA mode when $G_3$ is now in the IA
    mode. The next IT
    operation only occurs after all the guess inputs complete their individual
    IA operations. This will be followed by a new round of $\mathrm{IA}\rightarrow\mathrm{UA}\rightarrow\mathrm{IT}$
    operations starting at $t^{(2)}_{\mathrm{IA},1}$ as
    shown in Fig. 4(b).
\end{enumerate}
\begin{figure} [!htp]
\begin{center}
\epsfig{file=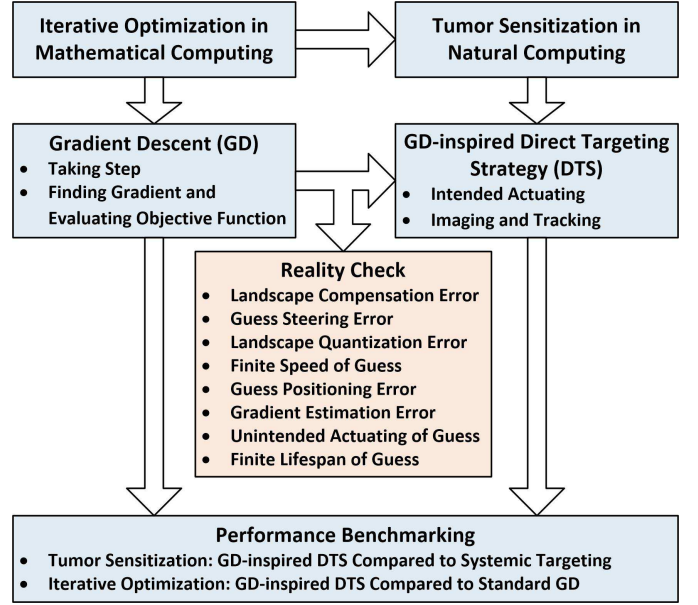,width=\linewidth}
\end{center}
    \caption{Mapping from iterative optimization in mathematical computing to
    tumor sensitization in natural computing.}
    \label{fig:Fig7}
\end{figure}

The mapping from an iterative optimization process in mathematical
computing to the aforementioned tumor sensitization process in
natural computing is illustrated in Fig. 7, which encompasses the
following procedures.
\begin{enumerate}
    \item \emph{General Mapping.} Formulate the nanoswimmers-assisted tumor
    sensitization in the perspective of natural computing
    as a stylized representation of the iterative optimization problem in mathematical computing.
    \item \emph{Specific Mapping.} Consider a specific
    iterative optimization algorithm $\mathcal{A}$, which is the
    GD in the current work, and map $\mathcal{A}$ onto the
    corresponding DTS $\mathcal{S}$. For example,
    the key operations in a standard GD include taking the step,
    finding the gradient, and evaluating the objective function. The
    first operation corresponds to the IA mode in the DTS and the
    last two operations are associated with the IT mode in the DTS.
    \item \emph{Reality Check.} Identify the key
    physical constraints associated with $\mathcal{S}$ when applied in a realistic
    \emph{in vivo} environment, compared to the original algorithm
    $\mathcal{A}$ when applied in an idealistic mathematical
    setting. For example, the imperfections in DTS include the
    landscape mismatch $\chi\left(\vec{x};G\right)$ in Eq. (2), the
    steering imperfection $\Delta\phi\left(t^{(1)}_{\mathrm{IA},1}\right)$ and the
    landscape quantization noise $\vec{q}_1\left(t^{(1)}_{\mathrm{IA},1}\right)$ in
    Eq. (3), the finite velocity of guess $v_1\left(t^{(1)}_{\mathrm{IA},1}\right)$ in Eq. (4),
    the positioning error $\Delta\vec{x}_1\left(t^{(1)}_{\mathrm{IT},1}+(k+1)\Delta t\right)$ and the gradient
    estimation inaccuracy in Eq. (5), the interference in guess update caused by
    UA, and the finite lifespan of guess inputs.
    \item \emph{Performance Benchmarking.} From the tumor sensitization
    perspective, we can evaluate the performance of the DTS $\mathcal{S}$ by
    comparing $\mathcal{S}$ to the ``brute-force'' systemic
    targeting without implementing any knowledge-aided targeting strategy. From the
    iterative optimization perspective, we can compare $\mathcal{S}$ to the standard
    algorithm $\mathcal{A}$. In this case, $\mathcal{S}$ is regarded as a degenerate
    form of $\mathcal{A}$.
\end{enumerate}

\subsection{Representative Objective Functions}
As the research is in its early stage, there is no widely-accepted,
quantitative model on any of the aforementioned biophysical
gradients in the existing literature other than some qualitative
observations made from experimental data. As an initial
investigation, three representative objective functions are
considered to evaluate the performance of the externally
manipulable, smart nanosystems for enhanced tumor sensitization as
shown in Fig. 8. The maximum value is normalized to $1$ and the
minimum value is $0$. The search domain is $-5~\mathrm{mm}\leq
x,y\leq5~\mathrm{mm}$. The landscapes are:
\begin{enumerate}
    \item \emph{Sphere Function (Bowl-shaped):}
    \begin{eqnarray} 
    \begin{split}
    &f(x,y)=\\
    &\left\{
    \begin{array}{ll}
    1,~~~~~~~~~~~\sqrt{x^2+y^2}\leq0.5~\mathrm{and}~\left(x,y\right)\in\mathbb{V}\\
    1-0.02\left(x^2+y^2\right),\\~~~~~~~~~~~~~~\sqrt{x^2+y^2}>0.5~\mathrm{and}~\left(x,y\right)\in\mathbb{V}.\\
    \end{array}\right.
    \end{split}
    \end{eqnarray}\begin{figure} [!htp]
\begin{center}
\epsfig{file=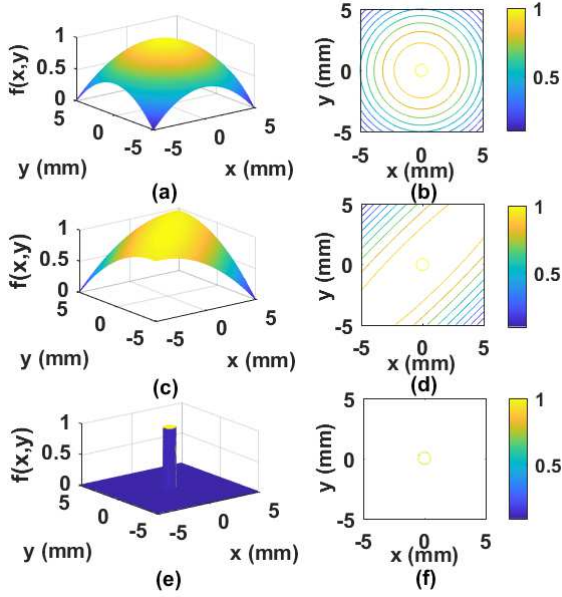,width=0.9\linewidth}
\end{center}
    \caption{Illustration of $f(x,y)$ for three representative objective functions:
    (a) Sphere function and (b) its contour plot;
    (c) Matyas function and (d) its contour plot; (e) Easom function
    and (f) its contour plot. For the objective $f(x,y)$, the maximum is normalized
    to 1 and the minimum value is 0.}
    \label{fig:Fig8}
\end{figure}
    \item \emph{Matyas Function (Plate-shaped):}
    \begin{eqnarray} 
    \begin{split}
    &f(x,y)=\\
    &\left\{
    \begin{array}{ll}
    1,~~~~~~~~~~~\sqrt{x^2+y^2}\leq0.5~\mathrm{and}~\left(x,y\right)\in\mathbb{V}\\
    1-0.01\left(x^2+y^2\right)+0.02xy,\\~~~~~~~~~~~~~~
    \sqrt{x^2+y^2}>0.5~\mathrm{and}~\left(x,y\right)\in\mathbb{V}.\\
    \end{array}\right.
    \end{split}
    \end{eqnarray}
    \item \emph{Easom function:}
    \begin{eqnarray} 
    \begin{split}
    &f\left(x,y\right)=\\
    &\left\{
    \begin{array}{ll}
    1,~~~~~~~~~~~\sqrt{x^2+y^2}\leq0.5~\mathrm{and}~\left(x,y\right)\in\mathbb{V}\\
    0.01+0.99\cos(3x)\cos(3y)\\
    ~~~~~~~~~~~~~~~\times\exp\left[-\left(9x^2+9y^2\right)\right],\\~~~~~~~~~~~~~~
    \sqrt{x^2+y^2}>0.5~\mathrm{and}~\left(x,y\right)\in\mathbb{V}.\\
    \end{array}\right.
    \end{split}
    \end{eqnarray}
    \end{enumerate}
The term $\mathbb{V}$ denotes the discrete vascular network as
illustrated in Fig. 3. As shown in Fig. 8(a)-(b), both the Sphere
and Matyas functions represent the situation that the tumor center,
denoted by a circle of radius $0.5~\mathrm{mm}$ located at the
origin, is associated with the region having the highest values of
$f(x,y)$. This may correspond to the largest (normalized) variation
of blood flow velocity due to tumor angiogenesis or the largest
(normalized) magnetic change induced by nanoswimmers due to fibrin
tropism in tumor tissues. Both the functions are convex and
quadratic. They have no local minimum except the global one. On the
other hand, the Easom function in Fig. 8(c) represents the situation
that $f(x,y)$ remains more or less unchanged across a large
surveillance region. The tumor center yields an abrupt increase of
$f(x,y)$. Intuitively, this may represent the worst-case direct
targeting scenario due to the lack of an externally observable
biophysical gradient.

In the absence of detailed information on the diameters of vessels,
the viscosity of blood, and the applied blood pressure for tumor
vessels and normal capillaries near the tumor, we simply imprint the
objective functions in (7)-(9) on the vascular network $\mathbb{V}$.
The blood inflow and outflow are assumed to be in the bottom left
and top right, respectively, where prescribed pressures are set.

\section{GD-inspired DTS}
The GD-inspired DTS starts with a generic guess $G_1$ located at
$\vec{x}_1$ at time instant $t^{(1)}_{\mathrm{IA},1}$, which
attempts to find a global maximum $f\left(\vec{x}^\ast\right)$.

\subsection{DTS for TDM1}
For TDM1 shown in Fig. 4(a), consider the sequence
$\vec{x}_1\left(t^{(1)}_{\mathrm{IT},1}\right),\vec{x}_1\left(t^{(2)}_{\mathrm{IT},1}\right),
\vec{x}_1\left(t^{(3)}_{\mathrm{IT},1}\right),\cdots$. In the
classical GD,
$\vec{x}_1\left(t^{(m)}_{\mathrm{IT},1}\right)=\vec{x}_1\left(t^{(m-1)}_{\mathrm{IT},1}\right)+\gamma_{m-1}\nabla
f\left(\vec{x}_1\left(t^{(m-1)}_{\mathrm{IT},1}\right)\right),m=2,3,\cdots$.
In this way, we have
$f\left(\vec{x}_1\left(t^{(1)}_{\mathrm{IT},1}\right)\right)\leq
f\left(\vec{x}_1\left(t^{(2)}_{\mathrm{IT},1}\right)\right)\leq
f\left(\vec{x}_1\left(t^{(3)}_{\mathrm{IT},1}\right)\right)\leq\cdots$,
so hopefully the sequence
$\vec{x}_1\left(t^{(m)}_{\mathrm{IT},1}\right)$ converges to the
desired global maximum. However, in the GD-inspired DTS, the
location updating is interrupted by multiple IT and UA processes as
depicted in Fig. 4(a). Hence, the position update is modified as
$\vec{x}_1\left(t^{(m)}_{\mathrm{IT},1}\right)=\vec{x}_1\left(t^{(m)}_{\mathrm{IA},1}\right)+\gamma_m\nabla
f\left(\vec{x}_1\left(t^{(m)}_{\mathrm{IA},1}\right)\right)$.

The gradient $\nabla
f\left(\vec{x}_1\left(t^{(m)}_{\mathrm{IA},1}\right)\right)$ is
estimated through the $N$ IT processes as illustrated in Fig. 4(a).
If the gradient does not change much over the duration of
$t^{(m)}_{\mathrm{IT},1}$ to $t^{(m+1)}_{\mathrm{IT},1}$, it can be
estimated as
\begin{equation} 
\nabla
f\left(\vec{x}_1\left(t^{(m)}_{\mathrm{IA},1}\right)\right)\approx\max_{n=1,2,\cdots,N}\left\{\nabla
f\left(\vec{x}_1\left(t^{(m)}_{\mathrm{IT},n}\right)\right)\right\}.
\end{equation}
Otherwise, only the last gradient estimate is used such that
\begin{equation} 
\nabla
f\left(\vec{x}_1\left(t^{(m)}_{\mathrm{IA},1}\right)\right)\approx\nabla
f\left(\vec{x}_1\left(t^{(m)}_{\mathrm{IT},N}\right)\right).
\end{equation}
Suppose that $f\left(\vec{x}\right)$ is convex and $\nabla
f\left(\vec{x}\right)$ is Lipschitz, the step size $\gamma_m$ can be
chosen to guarantee convergence to a global optimum by using the
Barzilai-Borwein method \cite{BB88}:
\begin{equation} 
\begin{split}
\gamma_m\approx&\frac{\left(\vec{x}_1\left(t^{(m)}_{\mathrm{IA},1}\right)-\vec{x}_1\left(t^{(m-1)}_{\mathrm{IA},1}\right)\right)^T}{\left\|\nabla
f\left(\vec{x}_1\left(t^{(m)}_{\mathrm{IA},1}\right)\right) -\nabla
f\left(\vec{x}_1\left(t^{(m-1)}_{\mathrm{IA},1}\right)\right)\right\|^2}\\
&\times\left[\nabla
f\left(\vec{x}_1\left(t^{(m)}_{\mathrm{IA},1}\right)\right) -\nabla
f\left(\vec{x}_1\left(t^{(m-1)}_{\mathrm{IA},1}\right)\right)\right].
\end{split}
\end{equation}

Note that the vessel network used in the simulation procedure is a
discontinuous two-dimensional grid as shown in Fig. 3; therefore the
position update follows the procedure described in Section III-B. As
the vessels run only parallel to the two coordinate axes, at each
junction the guess input can move in two possible directions, up and
right, as the flow is from bottom left to top right. The
Barzilai-Borwein condition in (12) is employed to determine the
duration of the $m^\mathrm{th}$ IA operation for $G_1$:
\begin{equation} 
t^{(m)}_{\mathrm{IT},1}-t^{(m)}_{\mathrm{IA},1}=\frac{\gamma_m\cos\phi_{m}+\gamma_{m}\sin\phi_{m}}{v_1\left(t^{(m)}_{\mathrm{IA},1}\right)},
\end{equation}
where $\phi_m$ is the angle of the gradient estimated at the
$m^\mathrm{th}$ cycle.

\subsection{DTS for TDM2}
For TDM2 shown in Fig. 4(b), similarly, consider the sequence
$\vec{x}_1\left(t^{(1)}_{\mathrm{IA},2}\right),\vec{x}_1\left(t^{(2)}_{\mathrm{IA},2}\right),
\vec{x}_1\left(t^{(3)}_{\mathrm{IA},2}\right),\cdots$. In the
classical GD,
$\vec{x}_1\left(t^{(m)}_{\mathrm{IA},2}\right)=\vec{x}_1\left(t^{(m-1)}_{\mathrm{IA},2}\right)+\gamma_{m-1}\nabla
f\left(\vec{x}_1\left(t^{(m-1)}_{\mathrm{IA},2}\right)\right),m=2,3,\cdots$,
to ensure that the sequence
$\vec{x}_1\left(t^{(m)}_{\mathrm{IA},2}\right)$ converges to the
desired global maximum. However, in the GD-inspired DTS, the
location updating is interrupted by multiple UA processes and one IT
process as depicted in Fig. 4(b). Hence, the position update is
expressed as
$\vec{x}_1\left(t^{(m)}_{\mathrm{IA},2}\right)=\vec{x}_1\left(t^{(m)}_{\mathrm{IA},1}\right)+\gamma_m\nabla
f\left(\vec{x}_1\left(t^{(m)}_{\mathrm{IA},1}\right)\right)$.

To ensure that such an arrangement does not favor IA processes that
are closer to the earlier IT operation resulted from more accurate
gradient estimation, the gradient change over the duration of
$t^{(m-1)}_{\mathrm{IA},1}$ to $t^{(m)}_{\mathrm{IA},1}$ should be
minimal, which is approximated by
\begin{equation} 
\nabla
f\left(\vec{x}_1\left(t^{(m)}_{\mathrm{IA},1}\right)\right)\approx\nabla
f\left(\vec{x}_1\left(t^{(m-1)}_{\mathrm{IT}}\right)\right),
\end{equation}
where $\nabla
f\left(\vec{x}_1\left(t^{(m-1)}_{\mathrm{IT}}\right)\right)$ is the
gradient estimated during the $(m-1)^\mathrm{th}$ IT process.

Finally, due to the practical constraint of DTS, the initial
deployment region of the guess input is confined within a small
area, which is the injection site of nanoswimmers, instead of the
entire solution space. To further ensure that the guess input is
confined within the tissue region under screening, the replacement
strategy is implemented: a guess that travels outside the allowed
searching region is abandoned, which will degrade in the human body
without further maneuvering and tracking. A new guess is then
generated in the deployment area by injecting an aggregate of
nanoswimmers.

\section{Performance Analysis}
We use several numerical examples to evaluate the tumor
sensitization and targeting performance of the GD-inspired DTS,
which is compared to the brute-force search.

For the DTS, both the two protocols in Fig. 4 are considered where
two guess inputs are deployed for direct targeting. The durations of
IA and IT are set to be $10~\mathrm{s}$ and the number of
observation intervals during each IT operation (see also Fig. 6) is
$10$. The searching process is stopped if any guess reaches the
cancer center denoted by a circle of radius $0.5~\mathrm{mm}$ at the
origin as shown in Fig. 8. It is assumed that the other guess can be
guided to the tumor center upon successful detection if it has not
overshot the tumor location.
\begin{figure} [!htp]
\begin{center}
\epsfig{file=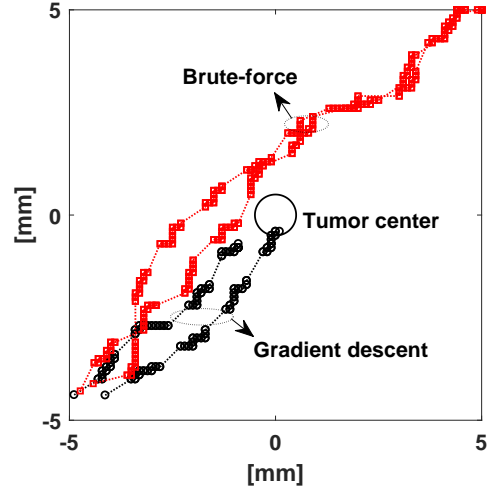,width=0.8\linewidth}
\end{center}
    \caption{Trajectories of guess inputs when TDM1 is applied: ``$\circ$'' - GD-inspired DTS, ``$\square$'' - brute-force search.}
    \label{fig:Fig9}
\end{figure}
\begin{figure} [!htp]
\begin{center}
\epsfig{file=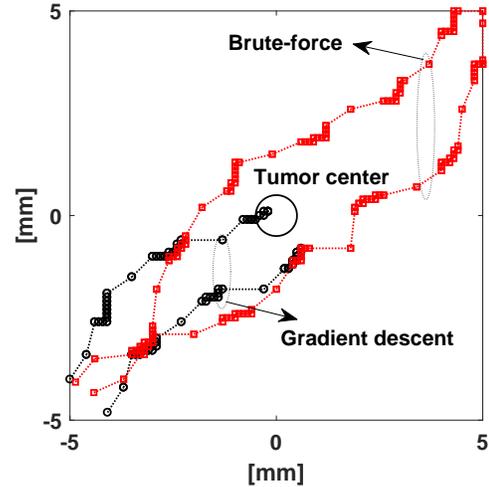,width=0.8\linewidth}
\end{center}
    \caption{Trajectories of guess inputs when TDM2 is applied: ``$\circ$'' - GD-inspired DTS, ``$\square$'' - brute-force search.}
    \label{fig:Fig10}
\end{figure}

For the brute-force search, each guess follows a random walk without
learning from the environment, which synthesizes the
contrast-enhanced medical imaging scenario where contrast agent
nanoparticles attempt to target a tumor by enumerating all possible
pathways in the vascular network. For consistency with the
GD-inspired DTS, the trajectories of two guess inputs are displayed.
Furthermore, the random drift and IT durations follow the two
protocols in Fig. 4, though the movement of these two guesses are
completely independent of each other. The searching process is
stopped if any guess reaches the cancer center as in the case of
DTS. However, the other guess will continue its random drift in the
absence of an external guidance.

The three objective functions presented in Section III-C are applied
to synthesize different levels of tumor sensitization difficulty.
The initial deployment region is $-5~\mathrm{mm}\leq
x,y\leq-4~\mathrm{mm}$. The speed of nanoswimmers is
$100~\mathrm{\mu m/s}$. The maximum search time allowed is
$200~\mathrm{s}$ and the total number of simulation runs is $1000$.
Two performance metrics are considered, the probability of cancer
detection $P_D$ and the percentage of contrast agent nanoparticles
delivered to the tumor site $\eta$.

Fig. 9 shows the typical trajectories of guess inputs for the
landscape of Sphere function when the TDM1 protocol is considered.
The symbols of ``$\circ$" and ``$\square$" denote the actual guess
footprints for the GD-inspired DTS and brute-force search,
respectively, and regions with clustered footprints correspond to
the IA mode. As can be seen from the figure, in the case of DTS the
movement of both guesses is coordinated by an external field towards
the maximum-gradient direction estimated in the IA mode. On the
other hand, the movement of two guesses is irregular and
uncorrelated for the brute-force search. The DTS successfully
detects the tumor center, whereas the brute-force search technique
fails to find the center of tissue malignancy even with multiple
guesses. Fig. 10 presents the guess trajectories when the TDM2
protocol is applied. The time interval between two consecutive IAs
for TDM2 is twice of the value for TDM1 because in the former case,
each guess takes turn to operate in the IA mode, followed by a
common IT mode. Hence, the gradient estimated during the IT
operation may be different from the actual gradient for the later IA
process, leading to more departing trajectories of the guesses as
depicted in Fig. 10. It is expected that this phenomenon would
result in deteriorating tumor sensitization and targeting
performance. Similar observations were made for the Matyas and Easom
landscapes.

Fig. 11 presents the histograms of search time for the three
objective functions when the TDM1 protocol is employed. The search
time of $200~\mathrm{s}$ (maximum value) indicates the situation
that none of the two guesses senses the tumor. It can be seen that
the GD-inspired DTS yields a detection ratio of $P_D=89.6\%$ for the
Sphere function (Fig. 11(a)), which is much higher than that for the
brute-force search ($P_D=58.1\%$, Fig. 11(b)). Furthermore, the DTS
has better performance in the Sphere landscape than the Matyas
($P_D=73.0\%$, Fig. 11(c)) and Easom ($P_D=59.8\%$, Fig. 11(e))
functions. This observation demonstrates the advantage of the
proposed biosensing-by-learning strategy over brute-force search and
the potential performance deterioration due to a more complex
landscape (i.e., plate-shaped Matyas function and gradientless Easom
function versus bowl-shaped Sphere function). In terms of the
targeting efficiency, the DTS achieves a much higher value of
$\eta=78.0\%$ compared to that for the Matyas $(\eta=61.6\%)$ and
Easom $(\eta=49.0\%)$ functions as well as the brute-force search
$(\eta\approx32\%)$.

Fig. 12 presents the histograms of search time for TDM2. Similar
observations to the trend in Fig. 11 can be made. Furthermore,
comparing Fig. 12 to Fig. 11 shows that, TDM1 yields higher
probabilities of tumor sensitization and larger percentages of drug
molecules delivered to the tumor than TDM2.

\begin{figure} [!htp]
\begin{center}
\epsfig{file=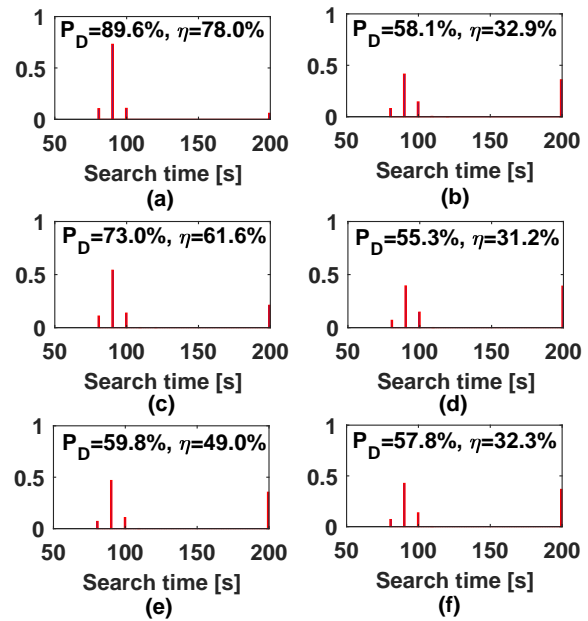,width=\linewidth}
\end{center}
    \caption{Histograms of search time when TDM1 is applied: (a) GD-inspired DTS and (b) brute-force search in a Sphere
    landscape; (c) GD-inspired DTS and (d) brute-force search in a Matyas landscape; (c) GD-inspired DTS
    and (d) brute-force search in an Easom landscape. Also shown are the respective detection ratios $P_D$
    and targeting efficiencies $\eta$.}
    \label{fig:Fig11}
\end{figure}
\begin{figure} [!htp]
\begin{center}
\epsfig{file=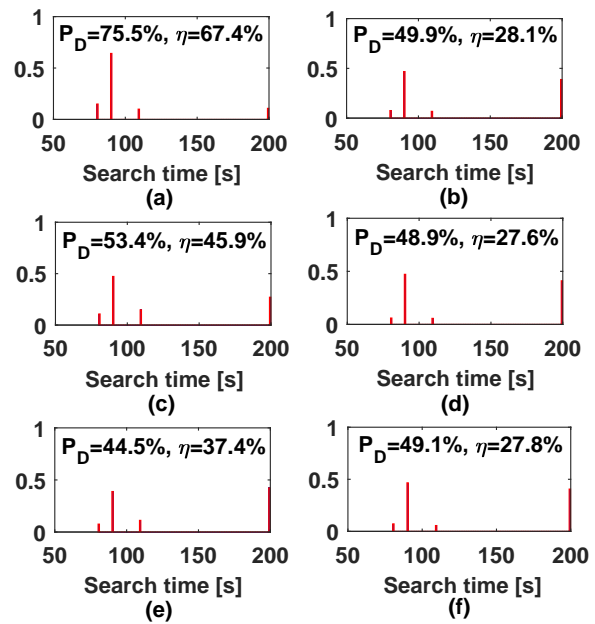,width=\linewidth}
\end{center}
    \caption{Histograms of search time when TDM2 is applied: (a) GD-inspired DTS and (b) brute-force search in a Sphere
    landscape; (c) GD-inspired DTS and (d) brute-force search in a Matyas landscape; (c) GD-inspired DTS
    and (d) brute-force search in an Easom landscape. Also shown are the respective detection ratios $P_D$
    and targeting efficiencies $\eta$.}
    \label{fig:Fig12}
\end{figure}

\section{Conclusion}
We have proposed a novel iterative-optimization-inspired DTS in
externally manipulable smart nanosystems, which exploits
tumor-triggered \emph{in vivo} biophysical gradients for ``guided''
direct targeting. We have demonstrated through computational
experiments that the proposed DTS can significantly improve the
probability of tumor sensitization and the accumulation of drug
nanoparticles in the tumor site by using the shortest possible
physiological routes and with minimal systemic exposure. We believe
that this work motivates a new paradigm directed toward smart
biosensing facilitated by externally controllable nanoswimmers.

Future work may include extension of the framework to DTS inspired
by multi-solution or multi-objective optimizations when there are
multiple tumors or different phenomena-of-interest in the tissue
region under surveillance. Moreover, it is important to examine
further the impact of nanoswimmer nonidealities, such as finite
lifespan, imprecise steering, and inaccurate tracking. Finally, the
proposed DTS and the objective functions used should be validated by
real experiments to justify further the clinical relevance of the
proposed strategy.

\bibliographystyle{ieeetr}
\bibliography{IEEEbib}

\end{document}